\documentstyle{article}
\input epsf
\catcode`\@=11
\long\def\@makefntext#1{ 
\protect\noindent \hbox to 3.2pt {\hskip-.9pt
$^{{\ninerm\@thefnmark}}$\hfil}#1\hfill} 
 
\def\thefootnote{\fnsymbol{footnote}}
 \def\@makefnmark{\hbox to 0pt{$^{\@thefnmark}$\hss}}  
	
\def\ps@myheadings{\let\@mkboth\@gobbletwo
\def\@oddhead{\hbox{} 
\rightmark\hfil\ninerm\thepage}
\def\@oddfoot{}\def\@evenhead{\ninerm\thepage\hfil 
\leftmark\hbox{}}\def\@evenfoot{}
\def\sectionmark##1{}\def\subsectionmark##1{}}
 
\newcommand{\be}{\begin{equation}}
\newcommand{\ee}{\end{equation}}
\newcommand{\ba}{\begin{eqnarray}}
\newcommand{\ea}{\end{eqnarray}}

\begin{document}

\def\abstracts#1{{
	\centering{\begin{minipage}{30pc}\tenrm\baselineskip=12pt\noindent
	\centerline{\tenrm ABSTRACT}\vspace{0.3cm}
	\parindent=0pt #1
	\end{minipage} }\par}}
 
 
\newcommand{\symbolfootnote}{\renewcommand{\thefootnote}
	{\fnsymbol{footnote}}}
\renewcommand{\thefootnote}{\fnsymbol{footnote}}
\newcommand{\alphfootnote}
	{\setcounter{footnote}{0}
	 \renewcommand{\thefootnote}{\sevenrm\alph{footnote}}}
 
\newcounter{sectionc}\newcounter{subsectionc}
\newcounter{subsubsectionc}
\renewcommand{\section}[1] {\vspace{0.6cm}\addtocounter{sectionc}{1}
\setcounter{subsectionc}{0}\setcounter{subsubsectionc}{0}\noindent
	{\bf\thesectionc. #1}\par\vspace{0.4cm}}
\renewcommand{\subsection}[1] {\vspace{0.6cm}
\addtocounter{subsectionc}{1}
	\setcounter{subsubsectionc}{0}\noindent
	{\it\thesectionc.\thesubsectionc. #1}\par\vspace{0.4cm}}
\renewcommand{\subsubsection}[1]
{\vspace{0.6cm}\addtocounter{subsubsectionc}{1}
	\noindent {\rm\thesectionc.\thesubsectionc.\thesubsubsectionc.
	#1}\par\vspace{0.4cm}}
\newcommand{\nonumsection}[1] {\vspace{0.6cm}\noindent{\bf #1}
	\par\vspace{0.4cm}}

\newcommand{\bibit}{\it}
\newcommand{\bibbf}{\bf}
\renewenvironment{thebibliography}[1]
	{\begin{list}{\arabic{enumi}.}
	{\usecounter{enumi}\setlength{\parsep}{0pt}
\setlength{\leftmargin 1.25cm}{\rightmargin 0pt}
	 \setlength{\itemsep}{0pt} \settowidth
	{\labelwidth}{#1.}\sloppy}}{\end{list}}
 
\topsep=0in\parsep=0in\itemsep=0in
\parindent=1.5pc
 
\newcounter{itemlistc}
\newcounter{romanlistc}
\newcounter{alphlistc}
\newcounter{arabiclistc}
\newenvironment{itemlist}
    	{\setcounter{itemlistc}{0}
	 \begin{list}{$\bullet$}
	{\usecounter{itemlistc}
	 \setlength{\parsep}{0pt}
	 \setlength{\itemsep}{0pt}}}{\end{list}}
 
\newenvironment{romanlist}
	{\setcounter{romanlistc}{0}
	 \begin{list}{$($\roman{romanlistc}$)$}
	{\usecounter{romanlistc}
	 \setlength{\parsep}{0pt}
	 \setlength{\itemsep}{0pt}}}{\end{list}}
 
\newenvironment{alphlist}
	{\setcounter{alphlistc}{0}
	 \begin{list}{$($\alph{alphlistc}$)$}
	{\usecounter{alphlistc}
	 \setlength{\parsep}{0pt}
	 \setlength{\itemsep}{0pt}}}{\end{list}}
 
\newenvironment{arabiclist}
	{\setcounter{arabiclistc}{0}
	 \begin{list}{\arabic{arabiclistc}}
	{\usecounter{arabiclistc}
	 \setlength{\parsep}{0pt}
	 \setlength{\itemsep}{0pt}}}{\end{list}}
 
\newcommand{\fcaption}[1]{
        \refstepcounter{figure}
        \setbox\@tempboxa = \hbox{\tenrm Fig.~\thefigure. #1}
        \ifdim \wd\@tempboxa > 6in
           {\begin{center}
        \parbox{6in}{\tenrm\baselineskip=12pt Fig.~\thefigure. #1 }
            \end{center}}
        \else
             {\begin{center}
             {\tenrm Fig.~\thefigure. #1}
              \end{center}}
        \fi}
 
\newcommand{\tcaption}[1]{
        \refstepcounter{table}
        \setbox\@tempboxa = \hbox{\tenrm Table~\thetable. #1}
        \ifdim \wd\@tempboxa > 6in
           {\begin{center}
        \parbox{6in}{\tenrm\baselineskip=12pt Table~\thetable. #1 }
            \end{center}}
        \else
             {\begin{center}
             {\tenrm Table~\thetable. #1}
              \end{center}}
        \fi}
 
\def\@citex[#1]#2{\if@filesw\immediate\write\@auxout
	{\string\citation{#2}}\fi
\def\@citea{}\@cite{\@for\@citeb:=#2\do
	{\@citea\def\@citea{,}\@ifundefined
	{b@\@citeb}{{\bf ?}\@warning
	{Citation `\@citeb' on page \thepage \space undefined}}
	{\csname b@\@citeb\endcsname}}}{#1}}
 
\newif\if@cghi
\def\cite{\@cghitrue\@ifnextchar [{\@tempswatrue
	\@citex}{\@tempswafalse\@citex[]}}
\def\citelow{\@cghifalse\@ifnextchar [{\@tempswatrue
	\@citex}{\@tempswafalse\@citex[]}}
\def\@cite#1#2{{$\null^{#1}$\if@tempswa\typeout
	{IJCGA warning: optional citation argument
	ignored: `#2'} \fi}}
\newcommand{\citeup}{\cite}
 
\def\fnm#1{$^{\mbox{\scriptsize #1}}$}
\def\fnt#1#2{\footnotetext{\kern-.3em
	{$^{\mbox{\sevenrm #1}}$}{#2}}}
 
\font\twelvebf=cmbx10 scaled\magstep 1
\font\twelverm=cmr10 scaled\magstep 1
\font\twelveit=cmti10 scaled\magstep 1
\font\elevenbfit=cmbxti10 scaled\magstephalf
\font\elevenbf=cmbx10 scaled\magstephalf
\font\elevenrm=cmr10 scaled\magstephalf
\font\elevenit=cmti10 scaled\magstephalf
\font\bfit=cmbxti10
\font\tenbf=cmbx10
\font\tenrm=cmr10
\font\tenit=cmti10
\font\ninebf=cmbx9
\font\ninerm=cmr9
\font\nineit=cmti9
\font\eightbf=cmbx8
\font\eightrm=cmr8
\font\eightit=cmti8

\begin{flushright}
CEBAF-TH-94-13
\end{flushright}

\begin{flushright}
hep-ph/9406237
\end{flushright}

\centerline{\tenbf  PION WAVE FUNCTION FROM  QCD
SUM RULES }
\baselineskip=16pt
\centerline{\tenbf WITH NONLOCAL CONDENSATES}
\baselineskip=13pt
\centerline{\tenit Talk  at the Workshop ``Continuous 
Advances in QCD'',}
\baselineskip=12pt
\centerline{\tenit February 18-20, TPI, University of 
Minnesota, Minneapolis}
\vspace{0.6cm}
\centerline{\tenrm  A.V.RADYUSHKIN\footnotemark}
\baselineskip=13pt
\centerline{\tenit Physics Department, Old Dominion
University}
\baselineskip=12pt
\centerline{\tenit  Norfolk, VA 23529, USA }
\baselineskip=12pt
\centerline{\tenit and}
\baselineskip=12pt
\centerline{\tenit Continuous Electron Beam Accelerator Facility}
\baselineskip=12pt
\centerline{\tenit  Newport News, VA 23606, USA  }
\vspace{0.6cm}

\footnotetext{Also at \em Laboratory of Theoretical Physics,
 Joint Institute for Nuclear Research,
141980 Dubna, Russian Federation }

\abstracts{We investigate a model QCD sum rule for the pion wave 
function
$\varphi_{\pi}(x)$
based on the non-diagonal correlator whose perturbative spectral 
density
vanishes and $\Phi(x,M^2)$,
the   theoretical  side of the sum rule, consists of
condensate contributions only.
We study the dependence of  $\Phi(x,M^2)$  on the Borel
parameter $M^2$ and observe that
 $\Phi(x,M^2)$ has a humpy form,
with the humps becoming   more and more pronounced when
$M^2$ increases. We demonstrate  that
this  phenomenon   reflects  just the oscillatory nature of the
higher states wave functions, while the
lowest state wave  function $\varphi_{\pi}(x)$
extracted from our QCD sum rule analysis,
  has no humps, is rather narrow and its shape is close
 to the  asymptotic form $\varphi_{\pi}^{as}(x) = 6x(1-x)$. }

\vspace{0.4cm}
\rm\baselineskip=14pt

\section{ QCD sum rules and pion wave function}

The pion wave function $\phi_{\pi}(x)$ is the basic object
in the perturbative QCD (pQCD) description of hard exclusive processes
involving the pion:  $\phi_{\pi}(x)$
is  the probability amplitude to find the pion in a state composed
of  its two valence quarks  carrying the fractions $xP$ and $(1-x)P $
of its large  longitudinal momentum  $P$.
More rigorously,  the pion wave function $\phi_{\pi}(x, \mu)$
can be defined as the function,  whose $N$-th moment  is given by the
matrix element of a  local operator with $N$ covariant derivatives
\cite{ar77,er80}:
\be
\{P^{\nu} P^{\nu_1} \ldots  P^{\nu_N} \} \int_0^1 \phi(x; \mu)\, x^N 
\, dx =
i^{N-1} \langle 0 | \bar d \gamma_5 \{ \gamma^{\nu} D^{\nu_1} \ldots 
D^{\nu_N} \}
u \, |\pi^+, P \rangle |_{\mu}
\label{eq:operators}
\ee
where $\{\ldots \}$ denotes the symmetric-traceless part of a tensor 
and
$\mu$ is the renormalization parameter for the composite operator $
\cal{O}_{\it N}$.
Instead of $x$,  it is  more convenient  sometimes to use the 
relative variable
$\xi$ defined by $x \equiv (1+\xi )/2$. In the limit of exact
 $u$-$d$ symmetry, the pion wave function is an even function
of $\xi$, $i.e.$, all  odd $\xi$-moments of $\phi(x; \mu)$ vanish.
 This    definition of the wave function implies
that its  integral
normalization is fixed by the matrix element of the axial current:
\be
\int_0^1 \phi(x; \mu)\, x^N \, dx  = f_{\pi},
\ee
where $f_{\pi} \approx 133 \, MeV$  is the pion decay constant.
In many cases, it is convenient to use
the normalized  wave function $\varphi_{\pi}(x; \mu)
\equiv  \phi_{\pi}(x; \mu)/f_{\pi}$
whose integral is simply 1.
The $\mu$-dependence of $\varphi_{\pi}(x; \mu)$ is governed by the 
evolution equation
which follows from the renormalization-group equation for the 
composite operators\cite{er80}.
In the $\mu \to \infty$ limit $\varphi_{\pi}(x, \mu)$  has
a simple and natural form \cite{er80} (see also\cite{bl78})
\be
\varphi_{\pi}(x; \mu \to  \infty )
\equiv \varphi_{\pi}^{as}(x) = 6  x(1-x).
  \label{eq:wfas}
\ee

However, one is usually interested in  the form of $\varphi_{\pi}(x; 
\mu )$
at low values of the renormalization  parameter
$\mu \sim 1 \, GeV$ relevant to experimentally accessible  situations.
This form is determined by non-perturbative QCD dynamics and, in 
principle,
it may strongly differ from the  asymptotic limit.
To calculate the pion wave function at
low values of the probing parameter $\mu$,
one should   take into account   non-perturbative
aspects of QCD. The closest to pQCD and one  of the most popular 
non-perturbative
approaches is provided  by   QCD sum  rules\cite{svz} which  
incorporate information
about the non-trivial structure of the QCD vacuum $via$ the
operator product expansion.

\begin{figure}[t]
  \vspace{-1.5cm}
  \epsfxsize=16cm
  \hspace*{-2cm} \epsffile{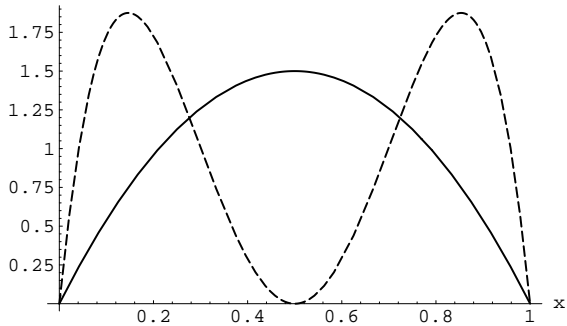}
  \vspace{-14cm}
 \centerline{\parbox{11cm}{\caption{\label{fig:1}
 Normalized pion wave function $ \varphi_\pi (x) $: asymptotic limit
(solid line) and CZ-model (dashed line).
   }}}
\end{figure}
 
The first application of  QCD sum rules
to the pion wave function $\phi_{\pi}(x)$ was the calculation of
the pion decay constant $f_{\pi}$, {\em i.e.,} the zeroth
moment of $\phi_{\pi}(x)$,
in  the pioneering  paper  by Shifman, Vainshtein and Zakharov
\cite{svz}
who  considered the correlator
of two axial currents and
 calculated $f_{\pi}$ within  5\% accuracy.   Next step was made by   
 Chernyak  and
A.Zhitnitsky\cite{cz82} who tried to construct the whole pion wave 
function  $\varphi_{\pi}(x)$
using information about its  lowest  moments calculated  from  the   
QCD sum rule
analysis for correlators of the axial current with the above mentioned
local operators  $\cal{O}_{\it N}$.
The model wave function chosen by CZ\cite{cz82}
\be
 \varphi^{CZ}_\pi (x)=30 \,x(1-x) (1-2x)^2 ,
                   \label{eq:wfcz}
\ee
has the moments $\langle\xi^2\rangle^{CZ} \approx 0.43$ and
$\langle\xi^4\rangle^{CZ} \approx 0.28$,
to be compared with $\langle\xi^2\rangle^{as}= 0.2$ and
$\langle\xi^4\rangle^{as} =3/35$ for the asymptotic wave function.
The large value of these moments dictates the
 characteristic double-humped shape of the CZ wave function: it has
maxima at $x \approx 0.15$
and $x \approx 0.85$ and a zero at the middlepoint  $x=0.5$ (see 
Fig.1).
Such a form for the  wave function of a lowest state
  looks rather strange.    Quantum-mechanics-based  intuition
would  rather suggest  that  the ground  state  wave function
has a shape like that of the asymptotic wave function: without
humps, nodes or zeros. One can also expect that all such 
peculiarities  should
 appear -- and  in  increasing number(!) --  for the wave functions of
radial  excitations.

Formally, the   reason for the   exotic shape of the CZ model wave 
function
  can be   traced to the
structure of the nonperturbative (condensate) terms  in
 their  sum rule.  Being written  directly for the wave function
$\varphi_{\pi}(x)$, it reads:
\ba
f_\pi^2\varphi_\pi(x) + (higher \ states)&=&\frac{3M^2}{2\pi^2} x(1-x)
  +\frac{\alpha_s\langle GG\rangle}{24\pi M^2}[\delta(x)+\delta(1-x)]
\nonumber \\
		&   + & \frac{8}{81}\frac{\pi\alpha_s\langle\bar
		   qq\rangle^2}{M^4}
\{11[\delta(x)+\delta(1-x)]+2[\delta^\prime(x)+\delta^\prime(1-x)]\}
                         \label{eq:wfsr}.
\ea
 
The  perturbative loop contribution in this sum rule
has a smooth behavior  coinciding with the
asymptotic $x(1-x)$ shape. On the other hand,
the condensate terms are strongly peaked at the end-points $x=0$ and 
$x=1$,
$i.e.,$ in the regions where one of the quarks has zero momentum.
The standard ansatz   for  the higher  states is to model
them by  perturbative spectral density, which is  $\sim x(1-x)$
in this case. As a result, there remains only one state,  the pion, 
whose wave function
has to reflect the presence of the    condensate peaks at $x=0$ and 
$x=1$.
In other words, the CZ wave function looks like a compromise between
the smooth perturbative loop behavior $\sim x(1-x)$
and the condensate contributions forcing
strong enhancements  at $x=0$ and $x=1$.
 
Earlier\cite{nlwf}, we  argued that
taking the CZ sum rule (\ref{eq:wfsr}) at  face  value amounts
to assumption that  vacuum quarks have zero momentum,
which is an approximation with a limited applicability range.
In general, one would  expect that
vacuum quarks  have a smooth  distribution in momentum, and, hence,
a $\delta(x)$  term, say, should   be treated
only as the first term of an  expansion of a  smooth function
in a series over $\delta(x)$  and its derivatives.
If   the generating smooth  function  is not very narrow, then the 
condensate peaks
are not as drastic as the $\delta(x) + \delta(1-x)$-approximation,
and the impact of the condensate corrections
on   the pion wave function is much  milder\cite{nlwf}.

\section{Nondiagonal  correlator}

In  what follows, we would like to concentrate on another subtle point
of the standard QCD sum rule analysis of   hadronic wave functions,
namely, on the implicit assumption that
higher states can be modeled by (``are dual to'') a perturbative  
spectral density.
In fact, this assumption is in an obvious
conflict  with a standard quantum-mechanical
situation, when the ground state has a monotonous positive-definite 
wave function,
its first radial excitation has one zero, the second has two and
 higher state wave functions
become more and more oscillating.
To study this problem in its cleanest form,  it makes sense to
 analize   a  sum rule with  vanishing  perturbative density.
This can be easily arranged  by taking
  a non-diagonal correlator\cite{zhit94}, $e.g.,$
the  correlator of  the generic operator $\cal{O}_{\it N}$ with the
pseudoscalar current $\bar d \gamma_5 u$ rather than with the
axial current $\bar d \gamma_5 \gamma_{\nu} u  \equiv {\cal O}_0$.
Then, for massless quarks,  all
the perturbative terms ($i.e.,$ those corresponding to the
unity operator)  of the operator product expansion  are zero
because  an odd number of gamma-matrices would be involved in any 
trace,
and  only   condensate terms  appear on the theoretical side of the 
sum rule.
To the lowest order of $\alpha_s$, all these terms have a singular
behaviour  like $\delta(x), \delta'(x), \, etc.$

The correlator of $\bar d \gamma_5 u$ and  $\bar u \gamma_5 \gamma 
D^N d$
has  a remarkably simple structure in the  limiting    case $N=0$
both on the theoretical and phenomenological sides of the sum rule:  
\\
$a)$ there is only one particle -- the pion -- which  has
nonzero projections  on both the axial current $\bar  d \gamma_5 
\gamma_{\mu} u$
and the pseudoscalar current $\bar  d \gamma_5  u$ ; \\
$b)$ for massless quarks,  a single   operator -- the quark 
condensate -- survives in
the operator product expansion (a proof can be found
in the SVZ paper\cite{svz}).   Two other,  formally  allowed
terms $\langle \bar q D^2 q \rangle$ and  $\langle \bar q (\sigma G) 
q \rangle$
cancel each other (recall that $\langle \bar q D^2 q \rangle =
\frac1{2}\langle \bar q ig(\sigma G) q \rangle$).
This leads to the well-known PCAC relation
\be
\langle 0 | \bar d \gamma_5 u | \pi \rangle = \frac{i}{f_{\pi}}
(\langle \bar u u \rangle  + \langle \bar d d \rangle ) .
\ee

 The relevant QCD sum rule, in its  borelized form, with $M^2$
being the Borel parameter characterizing the exponential suppression
of the higher states contribution, looks  as  follows:
\be
\langle \xi^N \rangle_{\pi} + \langle \xi^N \rangle_{\pi'} e^{-m_{
\pi'}^2/M^2} +
 \langle \xi^N \rangle_{\pi^{\prime \prime}} e^{-m_{\pi^{\prime 
\prime}}^2/M^2}
+ (higher \  states) =
\frac{1+(-1)^N}{2}  +  O(1/M^2) ,
\ee
where $O(1/M^2)$ includes the power suppressed contributions due to  
higher
condensates $\langle \bar q D^2 q \rangle$,
$\langle \bar q (\sigma G) q \rangle$, $etc.$
These contributions vanish for $N=0$, $i.e.,$ their coefficients
contain factor $N$.
 
Furthermore,  since only  the pion term survives on the l.h.s. of 
this sum rule
in the specific case $N=0$,  we have:
\be
\langle \xi^{N=0} \rangle_{\pi} =1 \  , \  \  \
\langle \xi^{N=0} \rangle_{\pi'} = \langle \xi^{N=0} \rangle_{\pi''} 
= \ldots =0.
\ee
A simple observation is  that the (``axial'') wave functions of
the higher pseudoscalar mesons $\pi'$,  $\pi'' , \ldots$
must (!) have oscillations   to produce
zero total integrals.  In principle,
the pion wave function may have oscillations or humps as well,
but this is not mandatory.

As before, we  rewrite     the sum rule directly for the wave 
functions:
\be
\varphi_{\pi}(x) + \varphi_{\pi'}(x) e^{-m_{\pi'}^2/M^2} + \ldots  =
\frac{ \delta(x) + \delta(1-x)}{2} + a \langle \bar q D^2 q \rangle
\left \{\delta'(x) + \delta'(1-x) \right \} + \ldots \  .
\ee
Note, that higher condensates cannot have $ \delta(x) $ or $
\delta(1-x)$
coefficients, since all higher condensate terms must disappear
after one takes  the zeroth moment of this sum rule.
 
Our last  comment here is that  the smooth  functions $\varphi_{
\pi}(x), \varphi_{\pi'}(x),
\ldots $  on the l.h.s. of the sum rule can be  produced only  by an 
infinite
summation of  singular distributions $ \delta^{n}(x) , 
\delta^{n}(1-x)$
associated with the  local condensates.

\section{Nonlocal condensates}

In the coordinate representation, the
contribution of the simplest  diagram  is
given by the product of the
perturbative propagator $S(z) \sim (z^{\mu}\gamma_{\mu})/z^4$ and the 
nonlocal condensate
$\langle \bar q (0) q(z) \rangle $.  Next term (evaluated
 in the Fock-Schwinger gauge, which is the most convenient
for the QCD sum rule calculations)
is proportional to
$  \langle \bar q (0) (\sigma G(0)) q(z) \rangle \, (z^{\mu}\gamma_{
\mu})/z^2.$
Performing the Taylor expansion of the nonlocal condensates in $z^2$ 
produces the OPE
in terms of local condensates. The   resulting sum rule for
the wave function has the structure of an expansion over the delta 
functions
$ \delta (x) , \delta(1-x)$   and  their derivatives.

Our strategy\cite{nlwf} is to avoid the Taylor  expansion  to 
preserve the  smoothness
properties of the objects involved on the
theoretical side of the sum rule, $i.e.$,   keep  together
all terms generated by a particular nonlocal condensate.
As the next step,  we construct  model expressions for the nonlocal 
condensates
to see how the properties of the nonlocal condensates affect the
form of the pion wave function extracted from the relevant sum rule.

It is convenient to parametrize the  $z^2$-dependence of the
simplest bilocal  quark condensate
$ \langle\bar q(0)q(z)\rangle \equiv \langle\bar q(0)q(0)\rangle 
Q(z^2)$
with the help of a Laplace-type representation
\be
Q(z^2) =
\int_{0}^{\infty} e^{s z^2/4}\, f(s)\, ds .
      \label{eq:qq}
\ee
The  spectral function $ f(s)$ may be called ``the distribution 
function of quarks
in the vacuum'' since its  $n$th moment is proportional to the
matrix  element of the local operator with  $D^2$ to $n$th power:
\be
\int_{0}^{\infty} s^N f(s)\,  ds \sim \frac{\langle\bar q (D^2)^N q 
\rangle}
{\langle\bar q  q \rangle}.
\ee
In particular, for the lowest two moments one has
\be
\int_{0}^{\infty}  f(s)\,  ds =1
\label{eq:norma}
\ee
and
\be
\int_{0}^{\infty} s f(s)\,  ds = \frac1{2} \frac{\langle\bar q (D^2)q 
\rangle}
{\langle\bar q  q \rangle} \equiv \frac{\lambda_q^2}{2},
\label{eq:lambda2}
\ee
with $\lambda_q^2$  having the meaning of  the average virtuality of  
vacuum quarks.

In a similar way, parametrizing the quark-gluon nonlocal condensate
$\langle \bar q(0) ig(\sigma G(0)) q(z) \rangle \equiv
\langle \bar q ig(\sigma G) q \rangle Q_1(z^2)$, one can introduce
the quark-gluon distribution function $f_1(s)$.
Since
\be
\langle\bar q D^2q\rangle = \frac1{2} \langle \bar q ig (\sigma G)q 
\rangle ,
\ee
there exist a relation between the zeroth moment of $f_1(s)$ and the 
first moment of
$f(s)$:
\be
m_0^2 \equiv \int_{0}^{\infty}  f_1(s)\,  ds  =  4 \int_{0}^{\infty}  
s f(s)\,  ds .
\label{eq:m02}
\ee
The standard QCD sum rule estimate\cite{belioffe} for $m_0^2$ is
$m_0^2 \simeq 0.8 \, GeV^2$; since $\lambda_q^2 = m_0^2/2$,
one can take $ \lambda_q^2 = 0.4 \, GeV^2$.

Constructing  models of  nonlocal condensates, one should satisfy 
also some other  constraints.
For instance, if one assumes that the  vacuum matrix element
$\langle\bar q (D^2)^{N_0} q \rangle$ exists, then $f(s)$ should 
vanish faster
than $1/s^{N_0+1}$ as $s \to \infty$.
So, if all such matrix elements exist, $f(s)$ must vanish faster than
any power of $1/s$ for large $s$.
As a possible choice, one may  impose that, at large $s$, the 
function $f(s)$
behaves like $ f(s) \sim e^{-s^2/\sigma^2}$ (Gaussian fall-off),
or $ f(s) \sim e^{-s/\sigma}$ (exponential fall-off), $etc.$
The opposite, small-$s$ limit of $f(s)$  is governed by the
large-$|z|$ properties  of the function $Q(z^2)$, $i.e.,$ by its
behaviour at large space separations or at large
values of the imaginary time variable $\tau = iz_0$.
The latter case can be easily assessed using
 the QCD sum rule for the heavy-light meson spectrum in
the  heavy quark effective theory (HQET).
In the HQET, the heavy quark has a trivial propagator
$S_Q(z) \sim  \delta^3({\bf z}) \theta(z_0)$
and, hence, the time dependence of the correlator of two heavy-light
currents is  determined by the  light quark propagator\cite{arhq}.
At large imaginary time $\tau$,  the correlator is dominated by the
lowest state contribution $\sim e^{-\tau \bar \Lambda}$
where $\bar \Lambda = (M_Q-m_Q)|_{m_Q\to \infty} $
is the lowest energy level of the mesons in HQET.
This means that $Q(z^2) \sim  e^{- |z| \bar \Lambda}$ for large 
Euclidean $z$
and $f(s) \sim e^{-\bar \Lambda^2/s}$  in the small-$s$ region.
Numerically, $\bar \Lambda$ is around 0.45 $GeV$.
 
Combining, in the simplest way, the  $e^{-\bar \Lambda^2/s}$  
dependence with,
say,   the  Gaussian fall-off at large $s$, we arrive at the ansatz
\be
f(s) = N e^{-\bar \Lambda^2/s - s^2/\sigma^2}
\ee
where $ \bar \Lambda^2 = 0.2 \, GeV^2$,
the normalization constant $N$ is fixed by eq.(\ref{eq:norma}) and the
$\sigma$-parameter is fixed by  eq.(\ref{eq:lambda2}), where for  the 
average
virtuality of vacuum quarks we take the
usual QCD sum rule value\cite{belioffe} $\lambda_q^2 \simeq 0.4 \, 
GeV^2$.

\begin{figure}[t]
  \vspace{-1.5cm}
  \epsfxsize=16cm
  \hspace*{-2cm} \epsffile{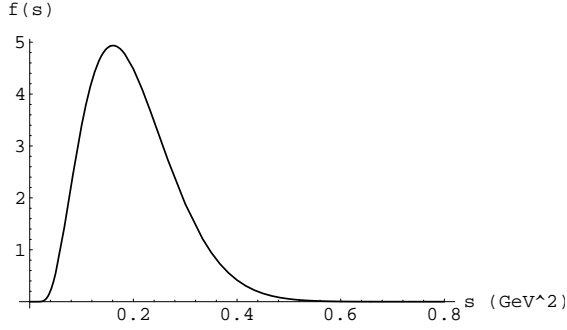}
  \vspace{-14cm}
 \centerline{\parbox{11cm}{\caption{\label{fig:2}
Our model for the vacuum distribution function  $f(s)$.
   }}}
\end{figure}

\section{Sum rule}
 
Now,  the sum rule for the pion wave function can be written as
\ba
\varphi_{\pi}(x) &+& \varphi_{\pi'}(x) e^{-m_{\pi'}^2/M^2} +
\varphi_{\pi''}(x) e^{-m_{\pi''}^2/M^2} + \ldots
\equiv  \Phi(x,M^2)  \\
&=& \left \{ \frac{M^2}{2}(1-x)  f(xM^2) + \frac1{8} f_1(xM^2)
 + F(x,M^2)  \right \}
+ (x \to 1-x) \nonumber
\ea
where $f, f_1,  \ldots$ are  the vacuum distribution functions
corresponding to the lowest nonlocal condensates $\langle \bar q q 
\rangle$
and $\langle \bar q \sigma G q \rangle$, respectively, and the 
function $F(x,M^2)$
is given by higher nonlocal condensates $\langle \bar q G G q 
\rangle$,
$\langle \bar q G G G q \rangle$,$etc.$
 
As discussed above, the sum rule reduces to an   extremely simple form
if one takes its zeroth moment:
\be
\int_0^1 \Phi(x,M^2) \, dx =1,
\label{eq:normalization}
\ee
with only the simplest quark condensate term contributing to ``1'' in 
the
r.h.s.
This imposes  the following    relations for the two lowest terms of 
the nonlocal
condensate expansion:
\ba
\int_{0}^{M^2}  f(s)\,  ds  =  1  ,  \nonumber \\
\int_{0}^{M^2}  f_1(s)\,  ds  = 4 \int_{0}^{M^2}  s f(s)\,  ds \,  .
\label{eq:relation}
\ea
The second relation reflects the cancellation between
$\langle \bar q \sigma G q \rangle$ and $\langle \bar q D^2 q \rangle$
terms of the local expansion.
For remaining condensates we have
\be
\int_{0}^{1}  F(x,M^2) \,  dx  =  0 .
\ee
By virtue of eqs.(\ref{eq:norma}, \ref{eq:m02}),
the first two  relations are  satisfied for  $M^2 = \infty$
and since, at large $s$, the distribution
functions are supposed to vanish
faster than any power of $1/s$,
the violation of the above finite-$M^2$ relations
also drops faster than any power of $1/M^2$ at large $M^2$.
This is consistent with the fact that
contributions  decreasing faster than any
power may be  missed by  the operator product expansion.
The practical lesson  is that  the Borel parameter
$M^2$  should  be taken in the region  where
the violation of the normalization
condition (\ref{eq:normalization})
is sufficiently small.

 To fix the sum rule, one  should specify  a model for  the
quark-gluon  nonlocal condensate and say something about the higher 
contributions
denoted by  $F(x,M^2)$.  As we will see, our  procedure of
extracting wave functions from the sum rule is perfectly linear,
in the sense that   each  nonlocal condensate term  on the
theoretical side of the sum rule produces an additive contribution
to all the  wave functions on its l.h.s.
  Hence,  one can split each wave function
into respective parts generated by $a)$ the lowest quark condensate,
$b)$ quark-gluon condensate, $c)$ next condensate, $etc.$ Then  one 
can
study separately the resulting  sum rules, each having only one type
of the nonlocal condensate on its  r.h.s.
Finally, one should add the contributions extracted from each of these
partial sum rules.
In fact, to  illustrate general features of the fitting procedure,
it is sufficient to analize the sum rule containing
the  lowest nonlocal condensate on its
theoretical side.
 However, since the two   lowest  nonlocal condensates are related by
eq.  (\ref{eq:relation}), these two terms
should be  better considered  together.

The structure of the  nonlocal quark-gluon condensate
is specified by the   relevant distribution function $f_1(s)$.
Information  about this function, in principle,   can be also obtained
from a (future)  study of the QCD sum rules in the heavy quark limit.
Lacking such information at the moment,
 we will assume  the  simplified ansatz
that $f_1(s)$ coincides with  the function  $f(s)$
governing the $z^2$-dependence of the simplest nonlocal quark 
condensate.
This assumption is not crucial and it does not  affect qualitative 
features of our analysis.

\section{Fitting  sum rule}

Now we can write down our model sum rule for the ``axial'' wave 
functions of the
pseudoscalar mesons:
\ba
\varphi_{\pi}(x) &+& \varphi_{\pi'}(x) e^{-m_{\pi'}^2/M^2}
+\varphi_{\pi''}(x) e^{-m_{\pi''}^2/M^2} + \ldots
\equiv  \Phi(x,M^2)  \\
&=& \frac{M^2}{2}\left (1-x+ \frac{\lambda_q^2}{2M^2} \right )  
f(xM^2) ,
+ (x \to 1-x) \nonumber
\ea
with the function $f(s)$ specified in the preceding section.
 For the $\pi'$-meson, we will take  the
experimental mass $m_{\pi'}^2 \simeq 1.7 \, GeV^2$.

It is evident from  this sum rule  that the function
$\Phi(x,M^2)$, $i.e.$, the weighted sum of all
wave functions is given by two  humps, which are moving as $M^2$ 
changes.
As $M^2$ increases, the  humps become narrower,   higher and move 
towards
 respective boundary points $x=0$ or $x=1$,
approaching the $\delta(x)$ or $ \delta(1-x)$ form  in the $M^2 \to 
\infty$ limit.
For $M^2=1 \, GeV^2$, $e.g.,$  the function  $\Phi(x,M^2)$ looks very 
much like the
Chernyak-Zhitnitsky wave function (see Fig.3). However, one should  
remember  that
$\Phi(x,M^2)$ is not just equal to the pion wave function: at large 
$M^2$ there might be
a large contamination from higher states.
Taking larger $M^2$, $e.g.$, $M^2 = 1.2 \, GeV^2$ produces even a 
wider function,
while decreasing  $M^2$  to $0.8 \, GeV^2$ produces a function with 
closer
and lower humps.

\begin{figure}[t]
  \vspace{-1.5cm}
  \epsfxsize=10cm
  \epsfysize=16cm
  \hspace{-1.5cm} \epsffile{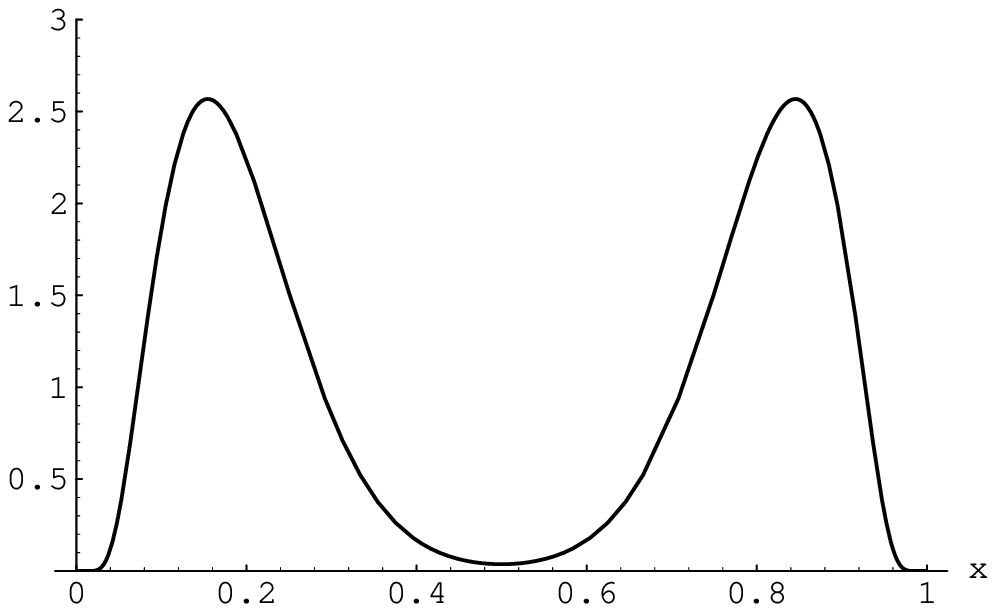}
 
  \vspace{-16cm}
  \epsfxsize=10cm
  \epsfysize=16cm
  \hspace{3.7cm} \epsffile{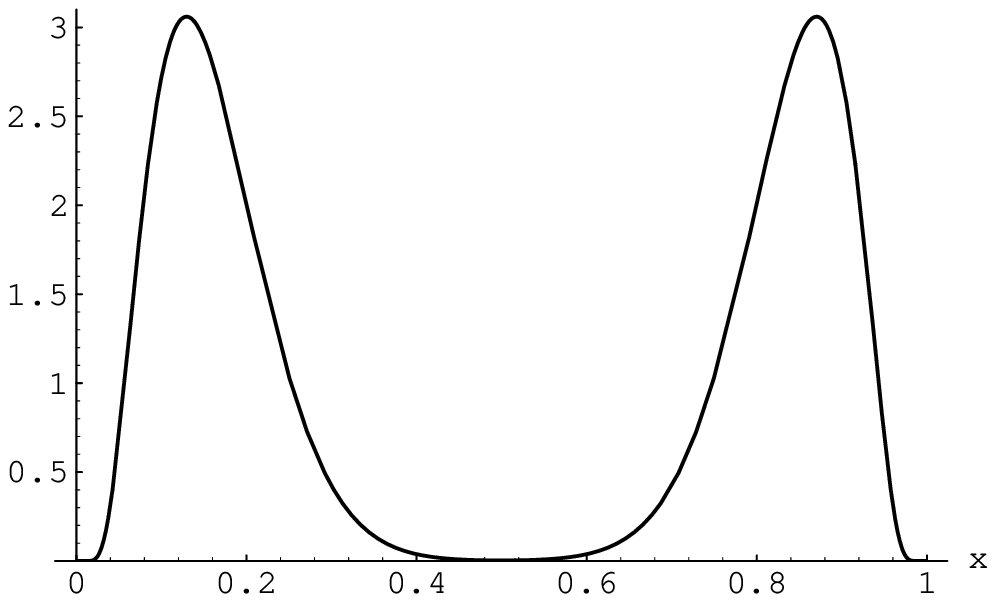}
 
  \vspace{-16cm}
  \epsfxsize=10cm
  \epsfysize=16cm
  \hspace*{8.9cm} \epsffile{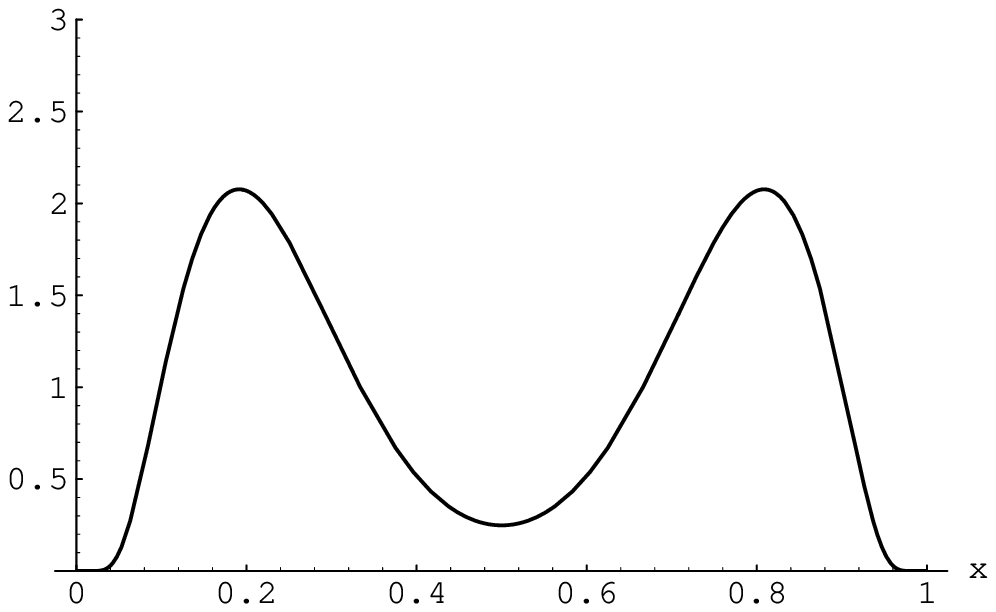}
  \vspace{-10.5cm}
 \centerline{\parbox{11cm}{\caption{\label{fig:3}
 Function $\Phi(x,M^2)$ \, for $M^2 = 1 \, GeV^2$  (left), \, 
$M^2=1.2 \, GeV^2 $ (middle)
 and  $M^2 = 0.8 \, GeV^2$ (right).
   }}}
\end{figure}

The lower  $M^2$, the  more pronounced is the dominance of the pion
in the total sum  $\Phi(x,M^2)$.
However, we cannot take too low $M^2$, because
the operator product expansion might fail.
Since the average virtuality $\lambda_q^2$ of the vacuum
quarks is $0.4 \, GeV^2$, it is definitely
unreasonable to go below the point $M^2 = 0.4 \, GeV^2$, because  our 
``large'' probing
virtuality $M^2$ should be larger than $\lambda_q^2$ -- otherwise one 
should expand
the correlator in $1/\lambda_q^2$ rather than in $1/M^2$.
Taking $M^2 = 0.4 \, GeV^2$, we observe that $\Phi(x,M^2)$  is very 
close to the
asymptotic  wave function of the pion (see Fig.4).
Assuming that  the total sum  $\Phi(x,M^2)$  at such low $M^2$ is 
completely dominated
by the pion,  we  have to conclude that our model for the nonlocal 
condensate sum rule
suggests that  the pion wave function  is rather close to its 
asymptotic form.
However, one should be more accurate here, since even the modest 
increases
of $M^2$ to $0.5 \, GeV^2$ or $0.6 \, GeV^2$ induce humps in $
\Phi(x,M^2)$ (see Fig.4).

\begin{figure}[h]
  \vspace{-1.5cm}
  \epsfxsize=10cm
  \epsfysize=16cm
  \hspace{-1.5cm} \epsffile{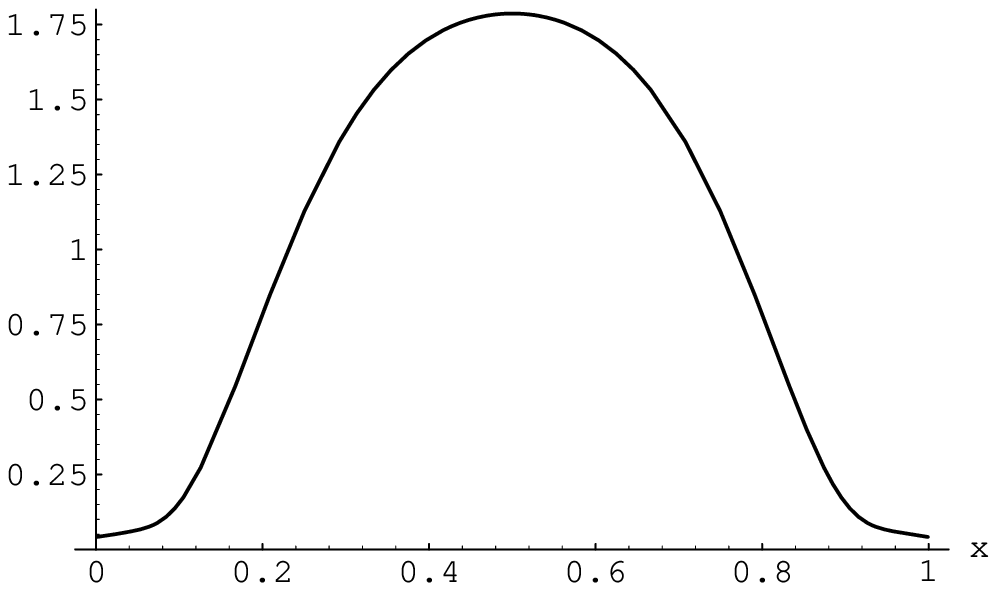}
 
  \vspace{-16cm}
  \epsfxsize=10cm
  \epsfysize=16cm
  \hspace{3.7cm} \epsffile{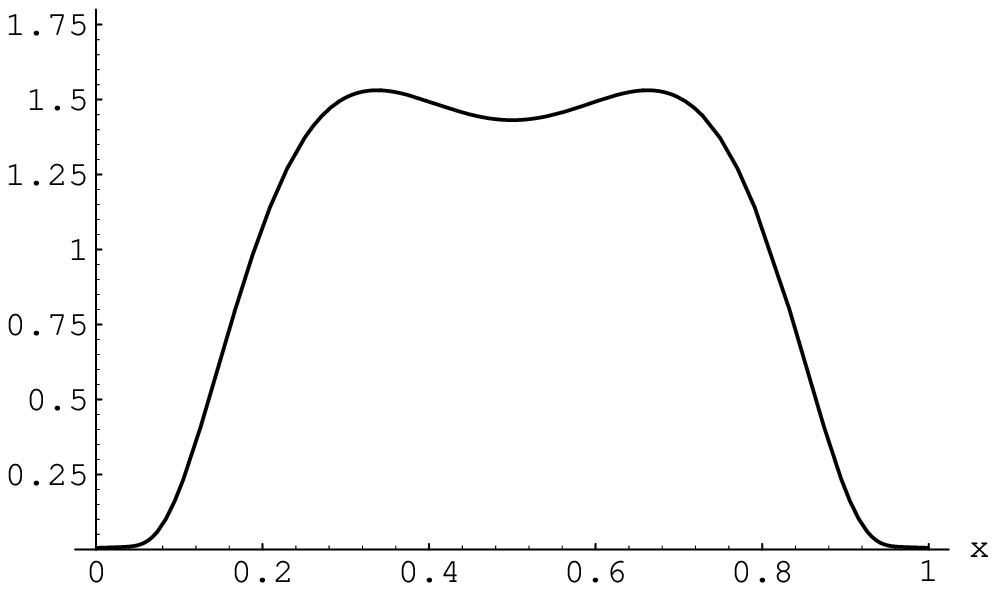}
 
  \vspace{-16cm}
  \epsfxsize=10cm
  \epsfysize=16cm
  \hspace*{8.9cm} \epsffile{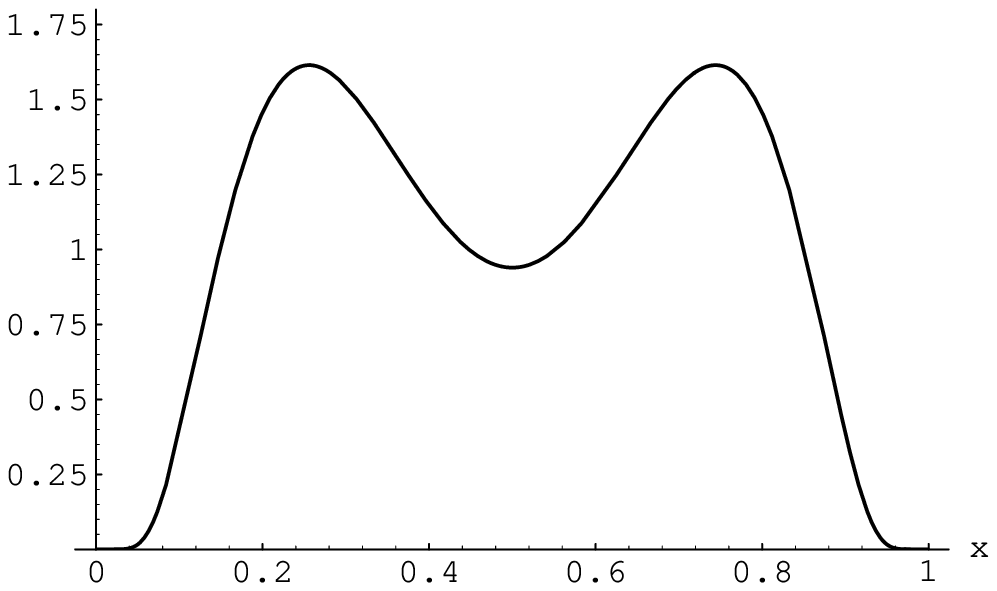}
  \vspace{-10.5cm}
 \centerline{\parbox{11cm}{\caption{\label{fig:4}
Function $\Phi(x,M^2)$ for $M^2= 0.4 \, GeV^2$ (left),   $ M^2=0.5  
\, GeV^2 $  (middle)
and $M^2= 0.6 \, GeV^2$ (right).
   }}}
\end{figure}

This means that the $\pi'$-contribution is visible at $M^2 \sim 0.5 
\, GeV^2$,
and one should better try to fit $\Phi(x,M^2)$ by two lowest states.
Taking two different but close values of $M^2$,
one can extract  the relevant wave functions,

\begin{figure}[t]
  \vspace{-1.5cm}
  \epsfxsize=10cm
  \epsfysize=16cm
  \hspace{-1.5cm} \epsffile{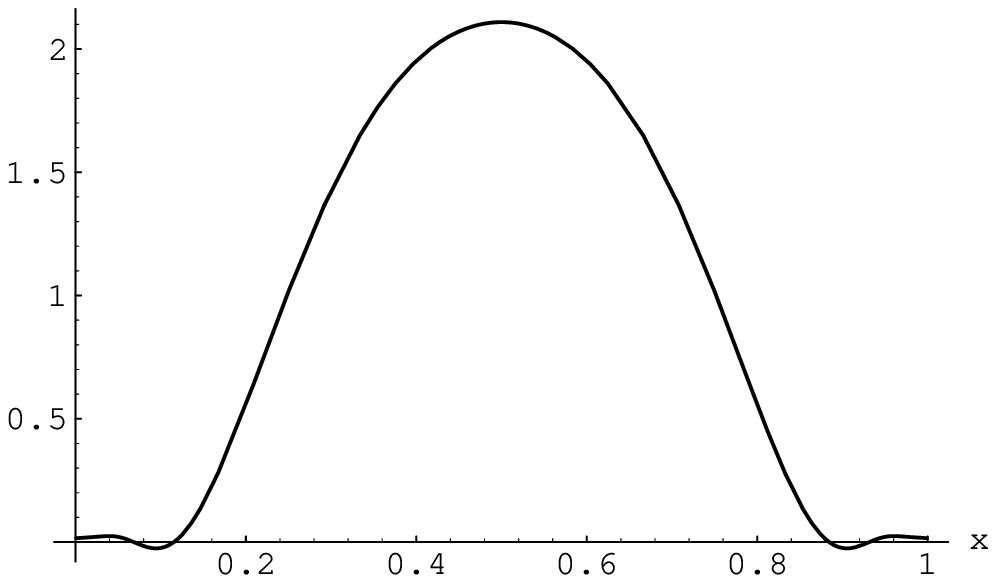}
 
  \vspace{-16cm}
  \epsfxsize=10cm
  \epsfysize=16cm
  \hspace{4cm} \epsffile{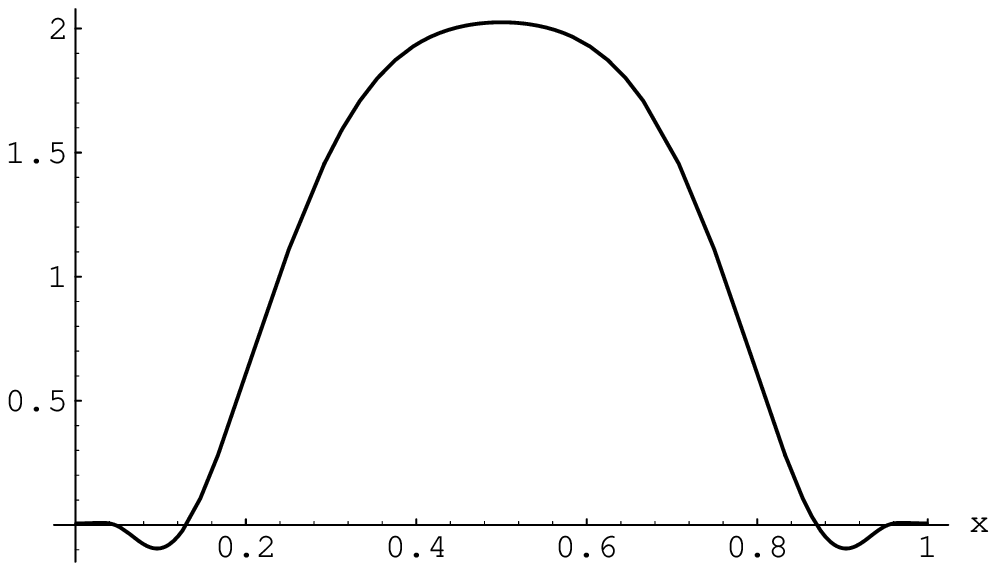}
 
  \vspace{-16cm}
  \epsfxsize=10cm
  \epsfysize=16cm
  \hspace*{9.5cm} \epsffile{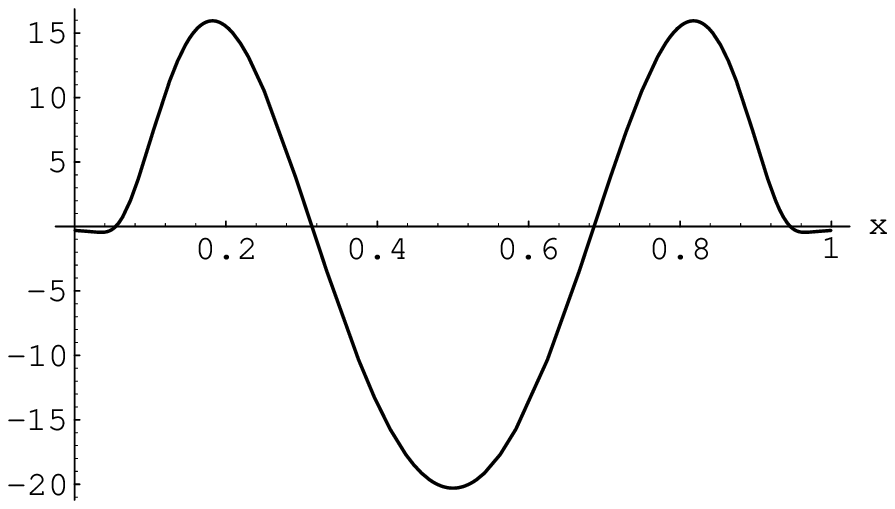}
  \vspace{-10.5cm}
 \centerline{\parbox{11cm}{\caption{\label{fig:5}
 Pion wave function extracted from  the
two-states fit of $\Phi(x,M^2)$ performed $a)$ \, at  $M_1^2= 0.5 \, 
GeV^2$
and  $M_2^2= 0.55 \, GeV^2$ (left) and $b)$ \, at  $M_1^2= 0.55 \, 
GeV^2$
and $M_2^2= 0.6 \, GeV^2$ (middle). $c)$ \, Wave function of the $
\pi'$-meson extracted from  the
two-states fit of $\Phi(x,M^2)$ performed at  $M_1^2= 0.5 \, GeV^2$
and $M_2^2= 0.55 \, GeV^2$ \,  (right).
   }}}
\end{figure}
 
\be
\varphi_{\pi}(x) \simeq
\frac{\Phi(x,M_1^2) e^{m_{\pi'}^2/M_1^2} - \Phi(x,M_2^2)
e^{m_{\pi'}^2/M_2^2}}{e^{m_{\pi'}^2/M_1^2} - e^{m_{\pi'}^2/M_2^2} } .
\ee
Choosing $M_1^2 =0.5 \, GeV^2$ and $M_1^2 =0.55 \, GeV^2$ we obtained 
the  curve
for $\varphi_{\pi}(x)$ shown in Fig.5 (left).
A very close result is obtained if one takes the pair $M_1^2 =0.55 \, 
GeV^2$ and $M_1^2 =0.60 \, GeV^2$
(see Fig.5, middle).
For definiteness, we will fix the pion wave function as that 
extracted from the
first pair $M_1^2 =0.5 \, GeV^2$ and $M_1^2 =0.55 \, GeV^2$.
The relevant $\pi'$ wave function is then shown in Fig.5 (right).
Note, that its maxima are by a factor of 10 higher than those of
$\varphi(x)$. However, this only means that the overall scale 
characterizing
the magnitude of matrix elements $\langle 0| \ldots | \pi' \rangle $
is  essentially larger  than $f_{\pi}$. But this is only natural in 
view
of large mass of the $\pi'$ particle.

 To estimate the contribution of the resonances higher than $\pi'$,
it is convenient to introduce the function
\be
\chi_{\pi'}(x,M^2) = \left ( \Phi(x,M^2) - \varphi_{\pi}(x) \right ) 
e^{m_{\pi'}^2/M^2} .
\ee
At low $M^2$, this function
is very close to   the $\pi'$  wave function $\varphi_{\pi'}(x)$
as determined from our  two-states fit.
In particular,  for the reference points
$M^2=0.5 \, GeV^2$ and  $M^2=0.55 \, GeV^2$, this function  simply
coincides with $\varphi_{\pi'}(x)$.
For larger $M^2$, however,  the
higher resonances  modify its form more and more  strongly.
This can be seen from Fig.6.   It is evident that the difference  
between
$\chi_{\pi'}(x,M^2)$ and  $ \varphi_{\pi'}(x) \equiv
\chi_{\pi'}(x,M^2 = 0.55 \, GeV^2)$ increases with $M^2$.
 One can guess that the increase of
$\chi_{\pi'}(x,M^2) - \varphi_{\pi'}(x)$  just
reflects the increasing contribution of the next resonance.
 Looking at the actual  curves for  the difference
$\chi_{\pi'}(x,M^2) - \varphi_{\pi'}(x)$ at  three values,
 $M^2= 0.8 \,GeV^2$ , $ M^2 = 1  \,GeV^2$  and $M^2 = 1.2 \, GeV^2$,
(see Fig.6 (right)), one can notice that, to good accuracy,
 the difference $\chi_{\pi'}(x,M^2) - \varphi_{\pi'}(x)$
 has  essentially the same shape for different $M^2$,
with absolute normalization governed by  an  $M^2$-dependent factor.

\begin{figure}[h]
  \vspace{-1.5cm}
  \epsfxsize=10cm
  \epsfysize=16cm
  \hspace{-1.5cm} \epsffile{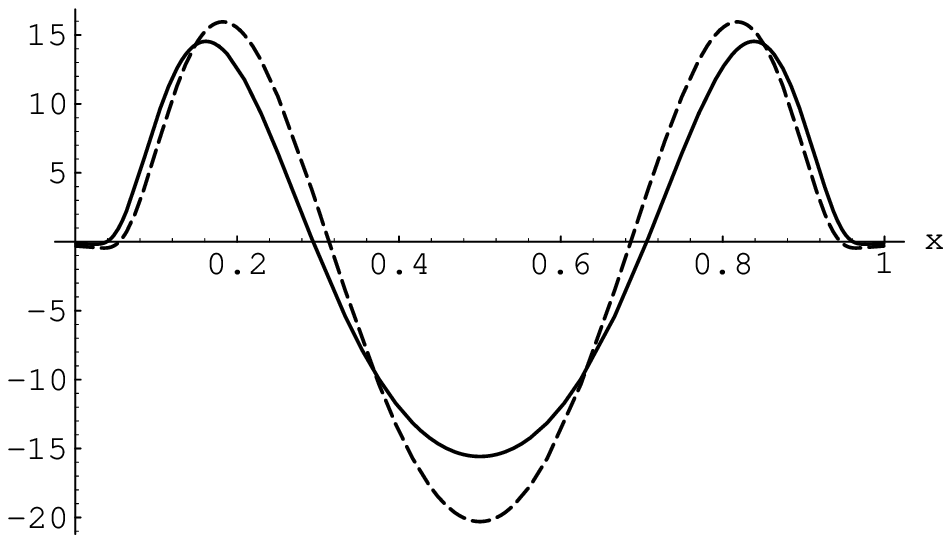}
 
  \vspace{-16cm}
  \epsfxsize=10cm
  \epsfysize=16cm
  \hspace{4cm} \epsffile{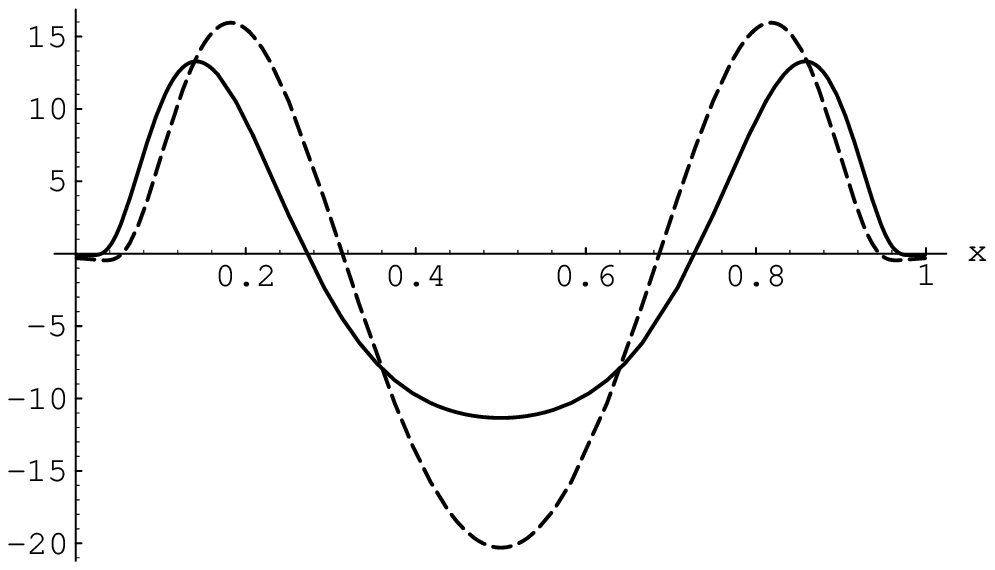}
 
  \vspace{-16cm}
  \epsfxsize=10cm
  \epsfysize=16cm
  \hspace*{9.5cm} \epsffile{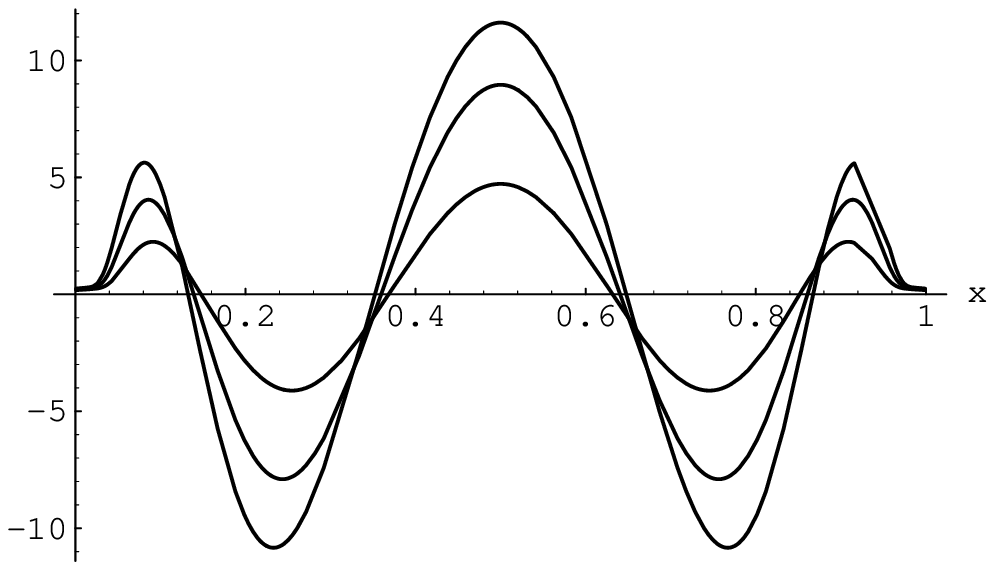}
  \vspace{-10.5cm}
 \centerline{\parbox{11cm}{\caption{\label{fig:6}
 Function $\chi_{\pi'}(x,M^2)$ for $M^2= 0.8 \, GeV^2$ (left),   and  
  $M^2 = 1 \, GeV^2$ (middle),
shown together with  $\varphi_{\pi'}(x)$ (dashed line).
Right picture shows the increase of the
difference $\chi_{\pi'}(x,M^2) - \varphi_{\pi'}(x)$
when the Borel parameter takes the values  $M^2= 0.8 , \, 1  $  and 
$1.2 \, GeV^2$.
   }}}
\end{figure}

Now, the question  is whether one can fit  the combination
\be
\Phi(x,M^2) - \varphi_{\pi}(x) - \varphi_{\pi'}(x) e^{-m_{
\pi'}^2/M^2} \equiv
(\chi_{\pi'}(x,M^2) - \varphi_{\pi'}(x))e^{-m_{\pi'}^2/M^2}
\ee
by the  next resonance  contribution
$
\varphi_{R}(x) e^{-m_{R}^2/M^2}  .
$
This means that  we should take the ratio
\be
{\chi_{\pi'}(x,M_1^2) - \varphi_{\pi'}(x)}\over{\chi_{\pi'}(x,M_2^2) 
- - \varphi_{\pi'}(x)}
\ee
and try to see whether it can be fitted by
$$
\frac{e^{m_{\pi'}^2/M_1^2-m_{R}^2/M_1^2}}{e^{m_{
\pi'}^2/M_2^2-m_{R}^2/M_2^2}}.
$$
This task can be reformulated as a  procedure  determining
  the mass of the third resonance from the relation
\be
m_R^2 - m_{\pi'}^2 =  \frac{M_1^2 M_2^2}{ M_1^2 - M_2^2 } \ln
\left [{\chi_{\pi'}(x,M_1^2) - \varphi_{\pi'}(x)}\over{\chi_{
\pi'}(x,M_2^2) - \varphi_{\pi'}(x)} \right ] .
\label{eq:log}
\ee

Again,  we take two pairs: $a) \, (M_1^2, \, M_2^2) = (1, \, 0.8) \, 
GeV^2$ and
$b) \, (M_1^2, \, M_2^2) = (1.2, \, 0.8) \, GeV^2$ and plot  the 
r.h.s. of eq.(\ref{eq:log})
for three $x$-regions (see Fig.7).
Because of the zeros of $\chi_{\pi'}(x,M_1^2) - \varphi_{\pi'}(x)$,
the curves are not as constant as one might wish.
Still, one can safely state that
 $m_R^2 - m_{\pi'}^2 = 2.5 \pm 0.5 \, GeV^2$ is a reasonable estimate.
For the mass itself, this gives
 $m_R^2 = 4.2 \pm 0.5  \, GeV^2$, which translates into a  rather 
narrow
prediction for the $\pi^{\prime \prime}$ mass:
$m_{\pi^{\prime \prime}} = 2.05 \pm 0.15 \, GeV$.

\begin{figure}[t]
  \vspace{-1.5cm}
  \epsfxsize=10cm
  \epsfysize=16cm
  \hspace{-1.5cm} \epsffile{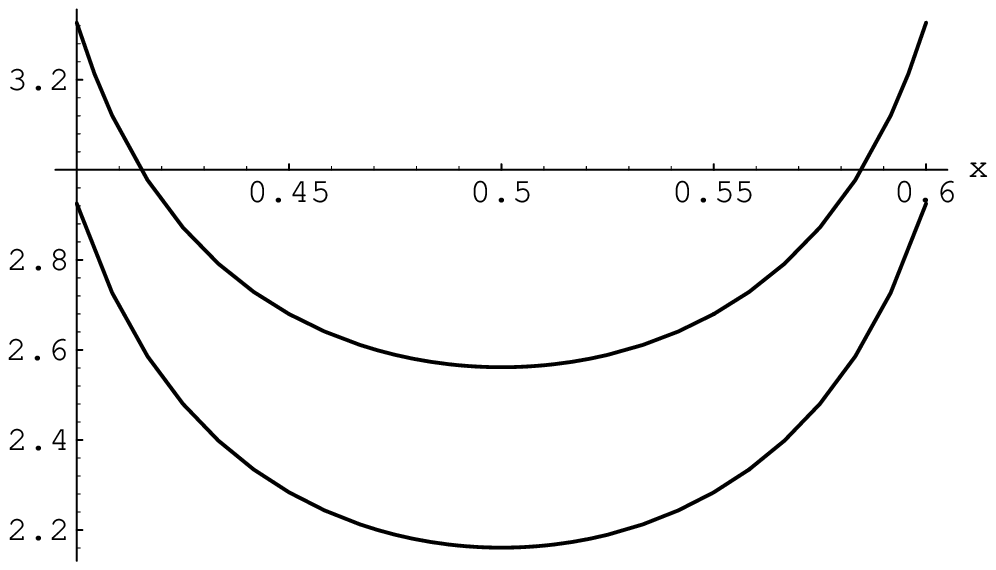}
 
  \vspace{-16cm}
  \epsfxsize=10cm
  \epsfysize=16cm
  \hspace{4cm} \epsffile{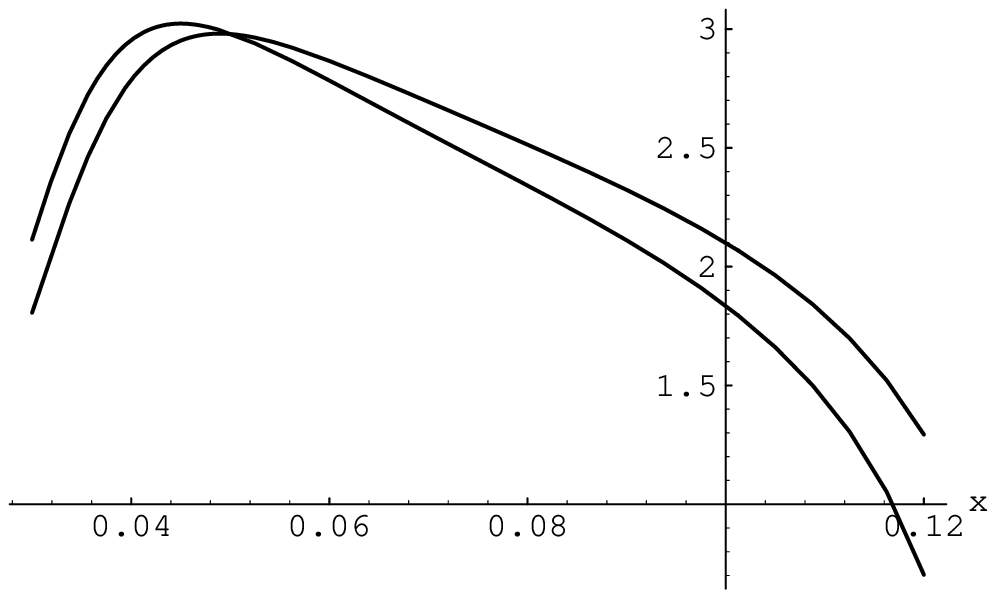}
 
  \vspace{-16cm}
  \epsfxsize=10cm
  \epsfysize=16cm
  \hspace*{9.5cm} \epsffile{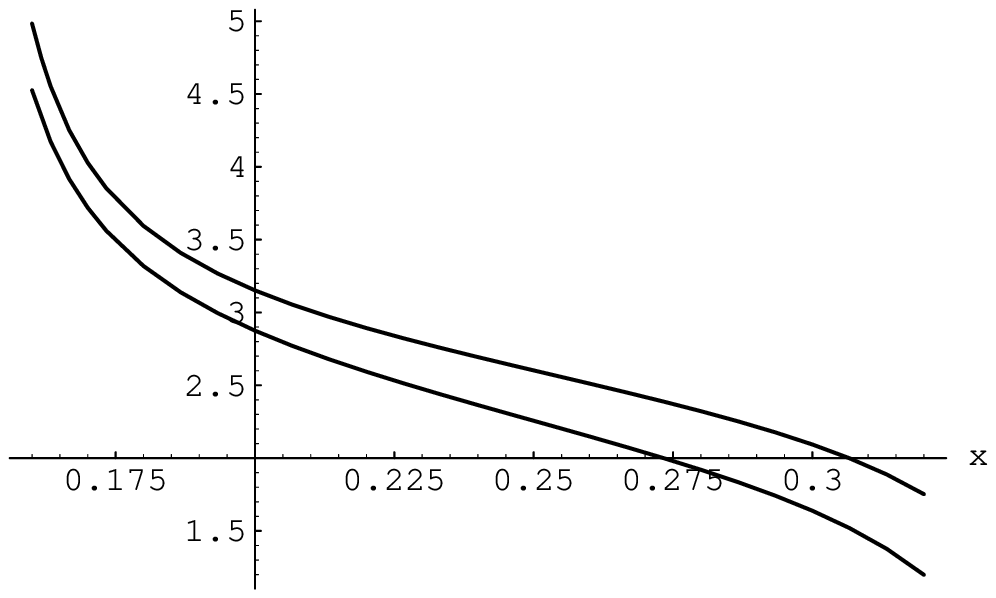}
  \vspace{-10.5cm}
 \centerline{\parbox{11cm}{\caption{\label{fig:7}
Mass difference $m_R^2 - m_{\pi'}^2$ calculated
{\it via} eq.(26) in essential $x$-regions.
   }}}
\end{figure}

\begin{figure}[h]
  \vspace{-1.5cm}
  \epsfxsize=10cm
  \epsfysize=16cm
  \hspace{-1.5cm} \epsffile{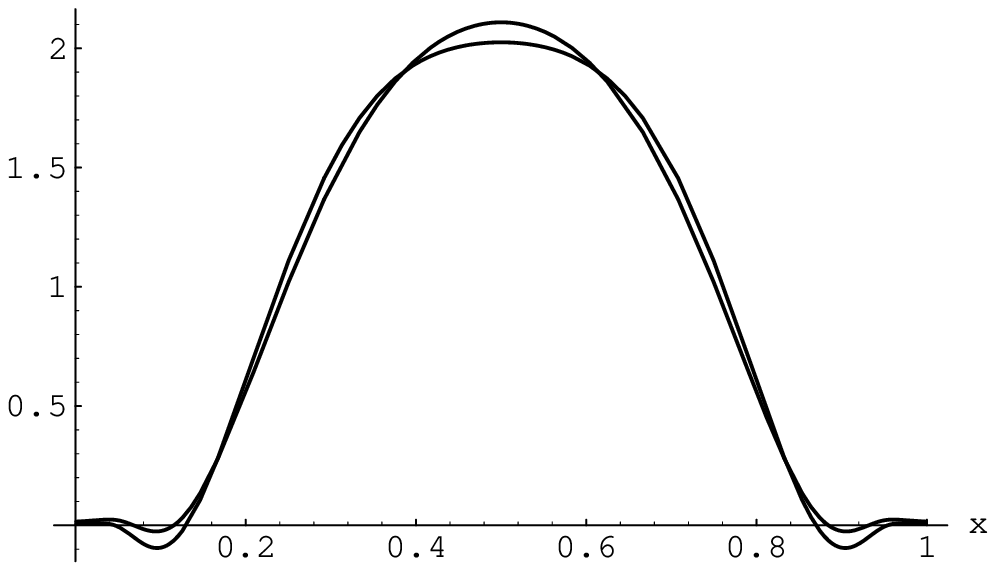}
 
  \vspace{-16cm}
  \epsfxsize=10cm
  \epsfysize=16cm
  \hspace{4cm} \epsffile{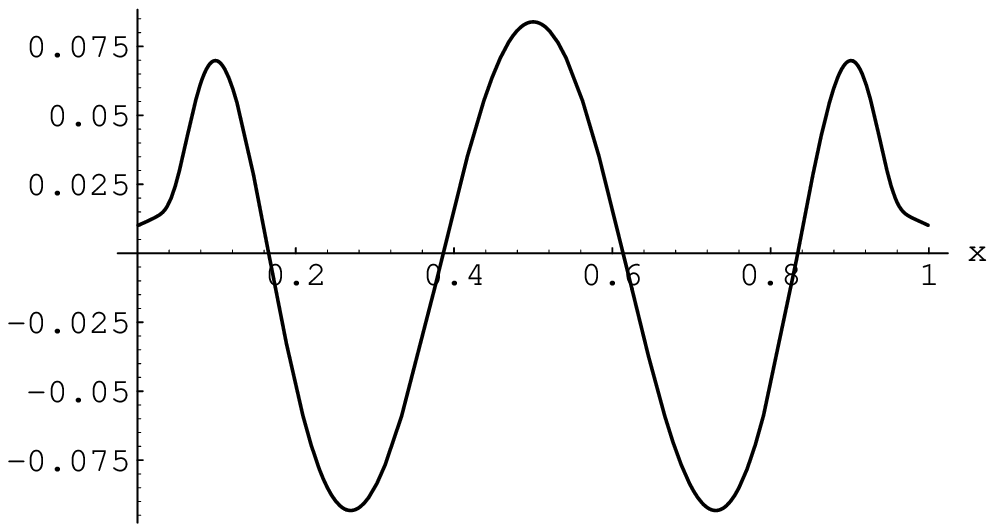}
  \vspace{-10.5cm}
 \centerline{\parbox{11cm}{\caption{\label{fig:8}
Pion wave functions from Fig.4 (left) and their difference (right).
   }}}
\end{figure}
 
Though  the third resonance is  rather massive,  it is not completely 
invisible
in the low-$M^2$ region. In particular,  looking at  the difference
between the two  pion wave functions  shown in Fig. 4 (recall that 
they were extracted from
two-states fits performed for different $M^2$ pairs), one can see that
the resulting curve has
 the shape  specific for the third resonance  (Fig.8).
 
\section{Summary and conclusions}

In this paper,  we considered a model sum rule for the pion wave 
function.
A specific feature of this sum rule is that  the  usual  perturbative 
contribution is
absent  altogether, and the theoretical side of the sum rule
is given by the condensate contributions only.
To represent the latter,  we  incorporated  nonlocal condensates.
As a result, $\Phi(x,M^2)$, the  weighted sum of wave functions 
related
to pion and its radial excitations,  was  given by a curve
generated by two humps  which were moving  to the end points $x=0$ 
and $x=1$,
raising in height with increasing
 Borel parameter $M^2$.  On the other hand, the relative weight
of the higher states increases when $M^2$  gets larger.
This  clearly indicates that the  peaks  observed in  $\Phi(x,M^2)$ 
for  $M^2 > 0.4 \, GeV^2$
reflect only the oscillatory nature of the wave functions related to 
the pion
excitations. Our explicit fits confirmed this expectation:
the pion wave function extracted from this sum rule has no humps and
is rather  narrow, despite all the humpy nature of $\Phi(x,M^2)$.
This result has evident implications for the CZ sum rule
based on a diagonal correlator:
the peaks in the end-point regions generated by  the
condensates contribution, reflect the oscillatory  components in the
higher states wave functions rather than the humpy
wave function of the ground state, the pion.

\section{Acknowledgements}
 
I am grateful to the organizers of the Workshop, especially
to M.Shifman and A.Vainshtein  for warm hospitality  at Minneapolis.
I  thank  I.Balitsky, V.Belyaev, I.Kogan,
G.Korchemsky  and A.Zhitnitsky for critical discussions.
 
This work was supported  by the US Department of Energy under 
contract DE-AC05-84ER40150.

\end{document}